\documentclass[twocolumn]{svjour3}  
\usepackage{multirow}
\usepackage{subfig}
\usepackage{graphicx}

\newcounter{subcopyrightbox@save}
\usepackage{caption}
\usepackage{color, url}
\usepackage{xspace} 
 \usepackage{balance}
 \usepackage{mathptmx} 
\smartqed
\newcommand{\eat}[1]{}

\newcommand{\myparatight}[1]{\smallskip\noindent{\bf {#1}:}~}

\begin{document}

\title{Reciprocal versus Parasocial Relationships in Online Social Networks}

\author{Neil Zhenqiang Gong \and
	Wenchang Xu\thanks{This work was done when W. Xu was a visiting student at UC Berkeley.}
}

\institute{ N. Z. Gong \at Computer Science Division, UC Berkeley \\
\email{neilz.gong@berkeley.edu}
\and
       W. Xu \at Department of Computer Science, Tsinghua University \\
       \email{wencxu@gmail.com}
}


\date{Received: date / Accepted: date}

\maketitle

\begin{abstract} 

Many online social networks are fundamentally \emph{directed}, i.e., they consist of both \emph{reciprocal} edges (i.e., edges that have already been linked back) and \emph{parasocial} edges (i.e., edges that haven't been linked back). Thus, {understanding the structures and evolutions of reciprocal edges and parasocial ones, exploring the factors that influence parasocial edges to become reciprocal ones, and predicting whether a parasocial edge will turn into a reciprocal one are basic research problems.}

However, there have been few systematic studies about such problems. In this paper, we bridge this gap using a novel large-scale Google+ dataset\footnote{Our Google+ dataset is available at \url{http://www.cs.berkeley.edu/~stevgong/dataset.html}} crawled by ourselves as well as one publicly available social network dataset.  First, we compare the structures and evolutions of reciprocal edges and those of parasocial edges. For instance, we find that reciprocal edges are more likely to connect users with similar degrees while parasocial edges are more likely to link \emph{ordinary} users (e.g., users with low degrees) and \emph{popular} users (e.g., celebrities).  However,  the impacts of reciprocal edges linking ordinary and popular users on the network structures increase slowly as the social networks evolve.
Second, we observe that factors including user behaviors, node attributes, and edge attributes all have significant impacts on the formation of reciprocal edges. Third, in contrast to previous studies that treat reciprocal edge prediction as either a supervised or a semi-supervised learning problem, we identify that reciprocal edge  prediction is better modeled as an outlier detection problem. Finally, we perform extensive evaluations with the two datasets, and we show that our proposal outperforms previous reciprocal edge prediction approaches. 

\end{abstract}

\section{Introduction}

Many online social networks (OSNs) such as Google+, Flickr, and Twitter are fundamentally \emph{directed}, i.e., that a user $u$ adds $v$ into his/her friend/followee list does not necessarily mean that $v$ also adds $u$ back. Thus, an OSN essentially consists of both \emph{reciprocal} edges (i.e., edges that have already been linked back) and \emph{parasocial} edges (i.e., edges that haven't been linked back). 

Therefore, understanding the interplays between reciprocal edges and parasocial ones in the evolution of OSNs and predicting whether a parasocial edge will turn into a reciprocal one (i.e., \emph{reciprocal edge prediction problem}) are basic research problems, the growing interest of which is highlighted by their importance in applications such as directed social network modeling, friend recommendation, information propagation, and network compression~\cite{Chierichetti09}.  

Reciprocal edge prediction is different from the classical \emph{link prediction} problem. First, previous work~\cite{Cheng11} has shown that features working well for link prediction are not the most effective ones for reciprocal edge prediction. Second, in the setting of reciprocal edge prediction, a parasocial edge already exists between two nodes, from which we can extract features. Third, they have very different domains. Specifically, the domain of the link prediction problem includes all non-existing edges in the network, which is huge. However, the domain of the reciprocal edge prediction problem consists of the parasocial edges, which is much smaller. The domain sizes have significant impact on the efficiency of the corresponding prediction algorithms.


Most existing measurement papers~\cite{Zhao12,Mislove08,Mislove07,Kwak10,Kumar06,Gong12-imc,Backstrom12,Ahn07,Seshadhri13} that study the structures and evolutions of OSNs either 1) tranform a directed OSN to an undirected one (e.g., keeping only the reciprocal edges) and then look into its structural properties (e.g., degree distribution, diameter, homophily, communities) and their evolutions or 2) look into basic structural properties (e.g., in/out degree distributions, reciprocity, triangles) of the original directed OSN and their evolutions.  Moreover, recent studies treated the reciprocal edge prediction problem as either a supervised learning problem~\cite{Cheng11} or a semi-supervised learning problem~\cite{Hopcroft11}. 

However, the differences between the structures and evolutions of reciprocal edges and those of parasocial edges are largely unexplored. It is also unclear that how user behaviors (e.g., tendencies of reciprocally linking back), node attributes (e.g., school, employer and major derived from users' profiles), and edge attributes (e.g., edge age) influence the formation of reciprocal edges. 

Moreover, treating reciprocal edge prediction as a supervised or a semi-supervised learning problem could incur a serious issue. Specifically, given a network snapshot, these approaches treat reciprocal edges as positive examples and sample some parasocial edges as negative examples to train models. Unfortunately, these sampled parasocial edges are also test examples  in the next snapshot. Moreover, some of them might have turned to be positive. As a result, the better the trained models are, the worse generalization performances they possibly achieve. In fact, in the setting of reciprocal edge prediction, we can only observe positive examples.

\myparatight{Our work} In this paper, we first compare the structures and evolutions of reciprocal edges and those of parasocial edges using a large-scale unique Google+ social network dataset that we crawled by ourselves and a Flickr social network dataset that was obtained from Mislove~\cite{Mislove08}. For instance, we find that reciprocal edges are more likely to link users with similar degrees while parasocial edges are more likely to connect ordinary users (i.e., users with low degrees) and popular users (e.g., celebrities).  However, the impacts of reciprocal edges linking ordinary and popular users on the network structures increase slowly as the social networks evolve. Second, we observe that user behaviors, node attributes, and edge attributes all have significant influences on the formation of reciprocal edges. For example, sharing common schools triples the probability of reciprocally linking back to a parasocial edge. These measurement results inform us the designs of features in the prediction of reciprocal edges. Third, in contrast to previous studies that treat reciprocal edge prediction as either a supervised or a semi-supervised learning problem, we model it as an outlier detection problem. Finally, we perform extensive evaluations with the Google+ and Flickr datasets, and we demonstrate that our proposal outperforms previous approaches.

To summarize, the key contributions of this work are:
\begin{itemize}
\item We perform the first study to compare the structures and evolutions of reciprocal edges and those of parasocial edges using a unique large-scale Google+ dataset crawled by ourselves and a publicly available large-scale dataset.
\item We find that user behaviors, node attributes, and edge attributes all have significant influences on the formation of reciprocal edges from parasocial edges, based on which we extract new features for predicting reciprocal edges.
\item We model reciprocal edge prediction as an outlier detection problem, in contrast to previous studies that treat reciprocal edge prediction as either a supervised or a semi-supervised learning problem. Moreover, we demonstrate that our proposal outperforms previous ones via extensively evaluating them using the two datasets.
\end{itemize}


\section{Notations and Datasets}
\label{sec:data}
We begin with the introduction of a few notations. Then, we describe our novel Google+ dataset and the publicly available Flickr dataset. The Google+ dataset represents an OSN's early stage while the Flickr dataset represents an OSN's steady stage, and thus they complement each other.

\begin{figure}
\centering
{\includegraphics[width=0.45 \textwidth, height=1.3in]{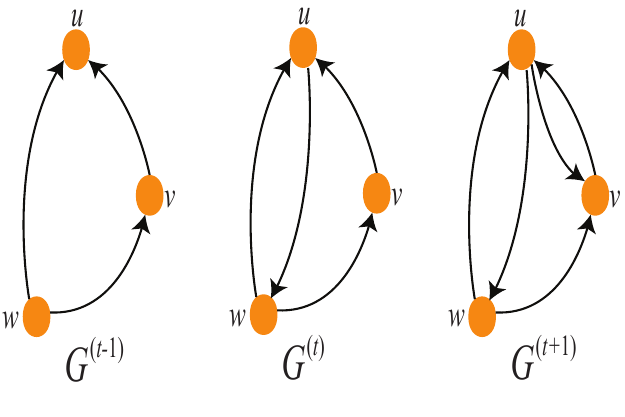}}
\caption{Illustration of friend requests, friend acceptances, parasocial edges, and reciprocal edges. Specifically, $(w,u)$, $(w,v)$, and $(v,u)$ are friend requests; $(u,w)$ is a friend acceptance; $(w,u)$ and $(u,w)$ are reciprocal edges; $(w,v)$ and $(v,u)$  are parasocial edges.}
\label{outlier}
\end{figure}

\subsection{Notations}
We denote a directed social network as $G=(V, E)$, where $V$ and $E$ are respectively the set of nodes and edges. We will elaborate how the nodes and edges are constructed when describing the datasets. Moreover,  the snapshot of $G$ at time $t$ is denoted as $G^{(t)}$. 


\begin{table*}[t]\renewcommand{\arraystretch}{1}
\centering
\caption{Statistics of a few basic network metrics of the largest snapshots of Google+ and Flickr.}
\addtolength{\tabcolsep}{-2.5pt}
\begin{tabular}{|c|c|c|c|c|c|c|c|} \hline
& \multirow{2}{*}{\#Nodes} & \multirow{2}{*}{\#Edges} & \multirow{2}{*}{Reciprocity} & \multicolumn{2}{|c|}{Assort. coeff.} & \multicolumn{2}{|c|}{Aver. clust. coeff.} \\ \cline{5-8} 
& & & & Parasocial & Reciprocal &Parasocial &Reciprocal \\ \hline
Google+ &{29,627,807} & {473,888,579} & {0.376} & -0.014 & 0.301 & 0.298 & 0.296 \\ \hline
Flickr & {2,302,925} & {33,140,018} & {0.451} & -0.003 & 0.127 &0.361 & 0.297 \\ \hline
\end{tabular}
\label{table:net_stat}
\end{table*} 

A directed edge $(u,v)$ is called as a \emph{friend request} if the reverse directed edge $(v,u)$ does not exist yet or appears after $(u,v)$\footnote{Note that some directed edges (e.g., in a Twitter follower network) don't really indicate ``friends''. Here, we denote them as friend requests for convenience. }, otherwise it's called as a \emph{friend acceptance}. Moreover, following the terminology in social science~\cite{Horton56}, we also classify edges be to \emph{parasocial} and \emph{reciprocal}. Specifically,  a directed edge $(u, v)$ is called parasocial if the reverse directed edge $(v,u)$ does not exist yet, otherwise it's called reciprocal.  According to the definitions, parasocial edges are friend requests that haven't been accepted yet. Figure~\ref{outlier} illustrates these concepts. For instance, in $G^{(t)}$, $(w,u)$, $(w,v)$ and $(v,u)$ are friend requests, $(u,w)$ is a friend acceptance, $(w,u)$ and $(u,w)$ are reciprocal edges and $(w,v)$ and $(v,u)$  are parasocial edges. 

We can undirect a directed social network $G=(V, E)$ in two ways, i.e., \emph{parasocial version} $G_{p}=(V_{p}, E_{p})$ and \emph{reciprocal version} $G_{r}=(V_{r}, E_{r})$, which satisfy that $V_{p}=V_{r}=V$, $E_{p}=\{(u,v)|(u,v) \in E\ \mathrm{or}\ (v,u) $ $\in E\}$ and $E_{r}=\{(u,v)|(u,v) \in E\ \mathrm{and}\ (v,u) \in E\}$. Intuitively, the reciprocal version consists of undirected reciprocal edges while the parasocial version includes both undirected reciprocal and parasocial edges.

For a  node $u$,  we denote its \emph{incoming neighbors} as
$\Gamma_{i}(u) = \{v|(v, u) \in E\}$ and \emph{indegree} as $d_i(u)=|\Gamma_{i}(u)|$,  \emph{outgoing neighbors} as $\Gamma_{o}(u) = \{v|(u, v) \in E\}$ and \emph{outdegree} as $d_o(u)=|\Gamma_{o}(u)|$,  \emph{parasocial neighbors} as $\Gamma_{p}(u)$ $=\Gamma_{i}(u)\cup \Gamma_{o}(u)$ and \emph{parasocial degree} as $d_p(u)=|\Gamma_{p}(u)|$, and \emph{reciprocal neighbors} as $\Gamma_{r}(u)=\Gamma_{i}(u)\cap \Gamma_{o}(u)$ and \emph{reciprocal degree} as $d_r(u)=|\Gamma_{r}(u)|$. 

Assume nodes also have binary attributes (e.g., Google Inc., Computer Science). For each binary attribute $a$, we denote its \emph{social neighbors} $\Gamma_{s}(a)$ as the set of nodes in $V$ that have the attribute $a$, and \emph{social degree} as $d_s(a) =|\Gamma_{s}(a)|$. Furthermore, we denote the set of attributes of a node $u$ as $\Gamma_a(u)$ and \emph{attribute degree} as $d_a(u) = |\Gamma_a(u)|$.

\subsection{Datasets}
\label{sec:dataset}
\myparatight{Google+} Google+ provides each user with an incoming friend list (i.e., ``have you in circles''), an outgoing friend list (i.e., ``in your circles'') and a profile page. We began to crawl daily snapshots of public Google+ social network structures and user profiles shortly after it was launched in late June, 2011; our dataset consists of 79 snapshots crawled from July 6 to October 11, 2011 (i.e., 98 days). The first snapshot was crawled by breadth-first search (without early stopping). On subsequent days, we expanded the social structure from the previous snapshot. For most snapshots, our crawl finished within one day as Google did not limit the crawling rate during that time. The 79 snapshots are denoted as $G^{(0)},G^{(1)},\cdots,G^{(97)}$, where superscripts are the normalized crawling dates of the snapshots. 
 
We take each user $u$ in Google+ as a node, and connect it to her outgoing friends via outgoing edges and incoming friends via incoming edges. Apart from the social structure, nodes also have attributes derived from users' profiles. We adopt four attribute types, i.e., \emph{School},
\emph{Major}, \emph{Employer} and \emph{City}. Specifically, we find all distinct schools, majors, employers and cities that appear in at least one crawled user
profile and use them as binary attributes. 

Gong et al.~\cite{Gong12-imc} roughly divided the evolution of Google+ into three phases: Phase I from day 1 to day 20, which corresponds to the early days of Google+
whose size increased dramatically; Phase II from day 21 to day 75,
during which Google+ went into a stabilized increasing phase; and
Phase III from day 76 to day 98, when Google+ was opened to public
(i.e., without requiring an invitation), resulting in a dramatic growth
again. We point out the three phases because reciprocal edges also evolve differently in them (see Section~\ref{sec:struct}).

\myparatight{Flickr} Flickr is a photo-sharing site based on
a social network, and it provides each user a friend list. This Flickr dataset, obtained from Mislove et al.~\cite{Mislove08},  has 102 snapshots crawled daily between
February 3rd, 2007 and May 18th, 2007.  We denote these snapshots as $G^{(0)}, G^{(1)}, \cdots, G^{(101)}$. This dataset represents a steady stage of Flickr since it was launched in 2002. We take each user as a node and connect it to its friends via outgoing edges.  Note that this dataset doesn't have node attributes.  


\myparatight{Summary} {Our Google+ dataset represents an OSN's \emph{early stage} while the Flickr dataset represents an OSN's \emph{steady stage}, making them complement each other.} Thus, as we will show in the following sections,   the two datasets have \emph{quantitatively} or even \emph{qualitatively} different structures and evolutions, and reciprocal prediction accuracy. 
 Table~\ref{table:net_stat} shows the statistics of a few basic network metrics of the largest snapshots of the Google+ and Flickr datasets. Reciprocity in the table is the fraction of friend requests that are already accepted.

\begin{figure*}[!t]
\centering
\subfloat[Google+]{\includegraphics[width=0.45 \textwidth, height=2in]{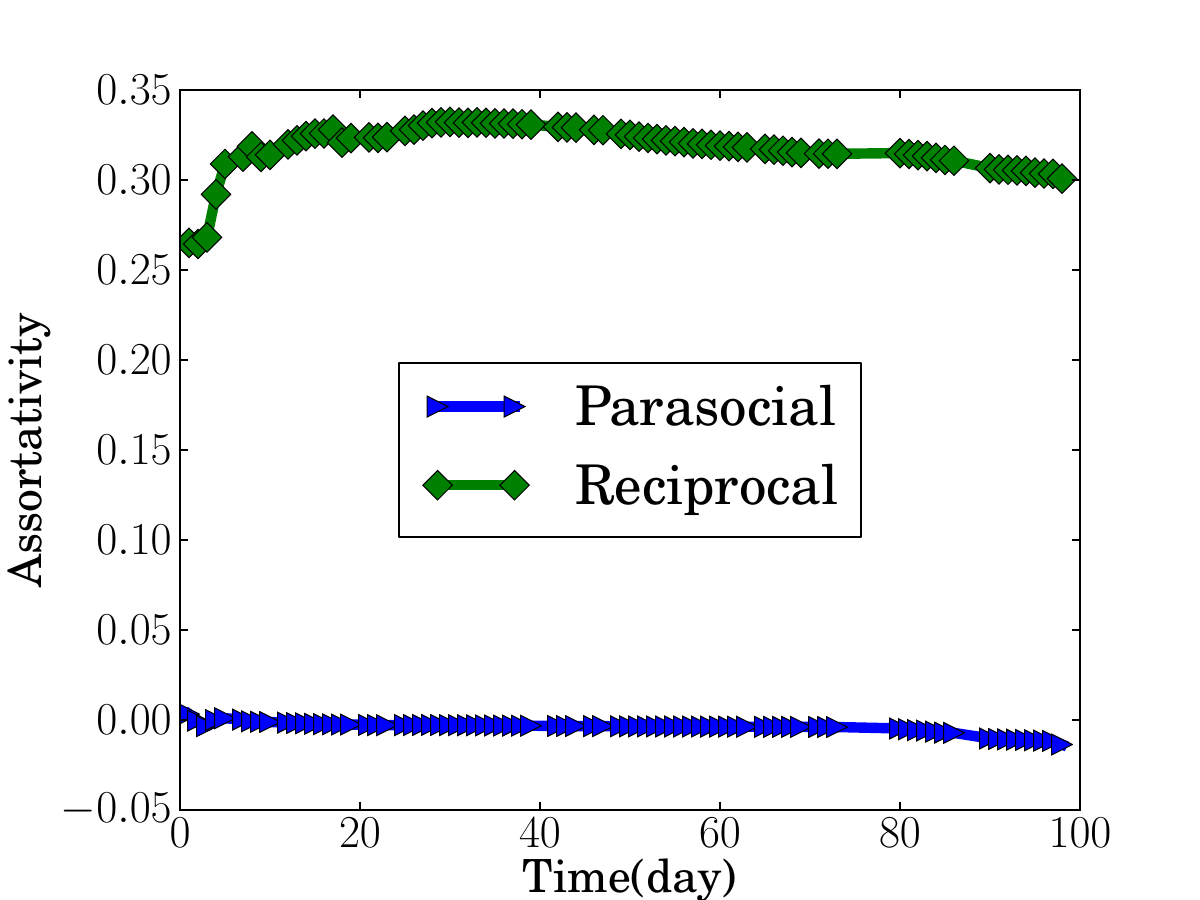} \label{gplus-assort-evo}}
\subfloat[Flickr]{\includegraphics[width=0.45 \textwidth, height=2in]{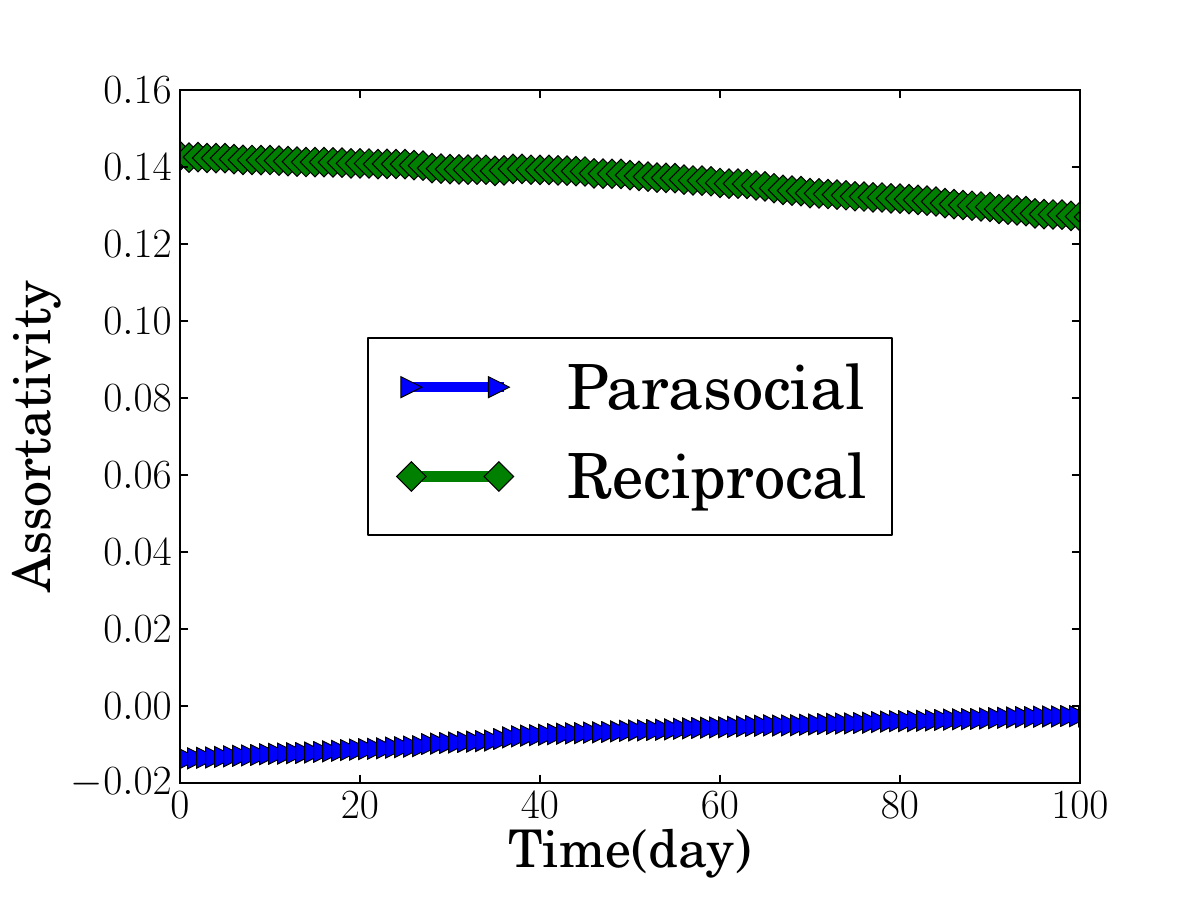} \label{flickr-assort-evo}}
\caption{Evolution of assortativity coefficients in (a) Google+ and (b) Flickr.}
\label{assort}
\end{figure*}

\section{Structure and Evolution}
\label{sec:struct}

In this section, we compare both global and local structures and their evolutions of reciprocal edges with those of  parasocial edges. Specifically, we explore the global one via studying degree homophily of the parasocial and reciprocal versions of the directed online social networks; we probe the local one via looking into the corresponding clustering coeffients.

\begin{figure*}[!t]
\centering
\subfloat[Google+]{\includegraphics[width=0.45 \textwidth, height=2in]{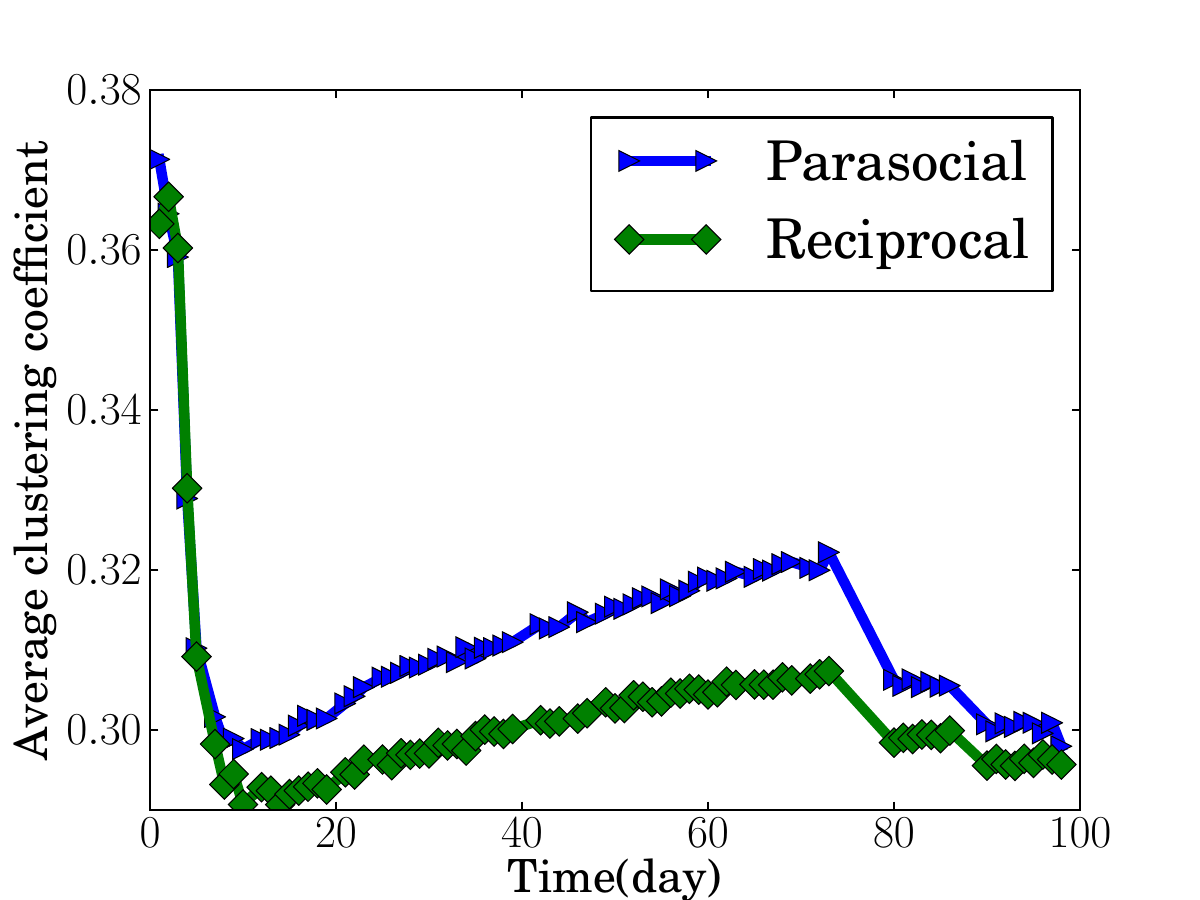} \label{gplus-clustering-evo}}
\subfloat[Flickr]{\includegraphics[width=0.45 \textwidth, height=2in]{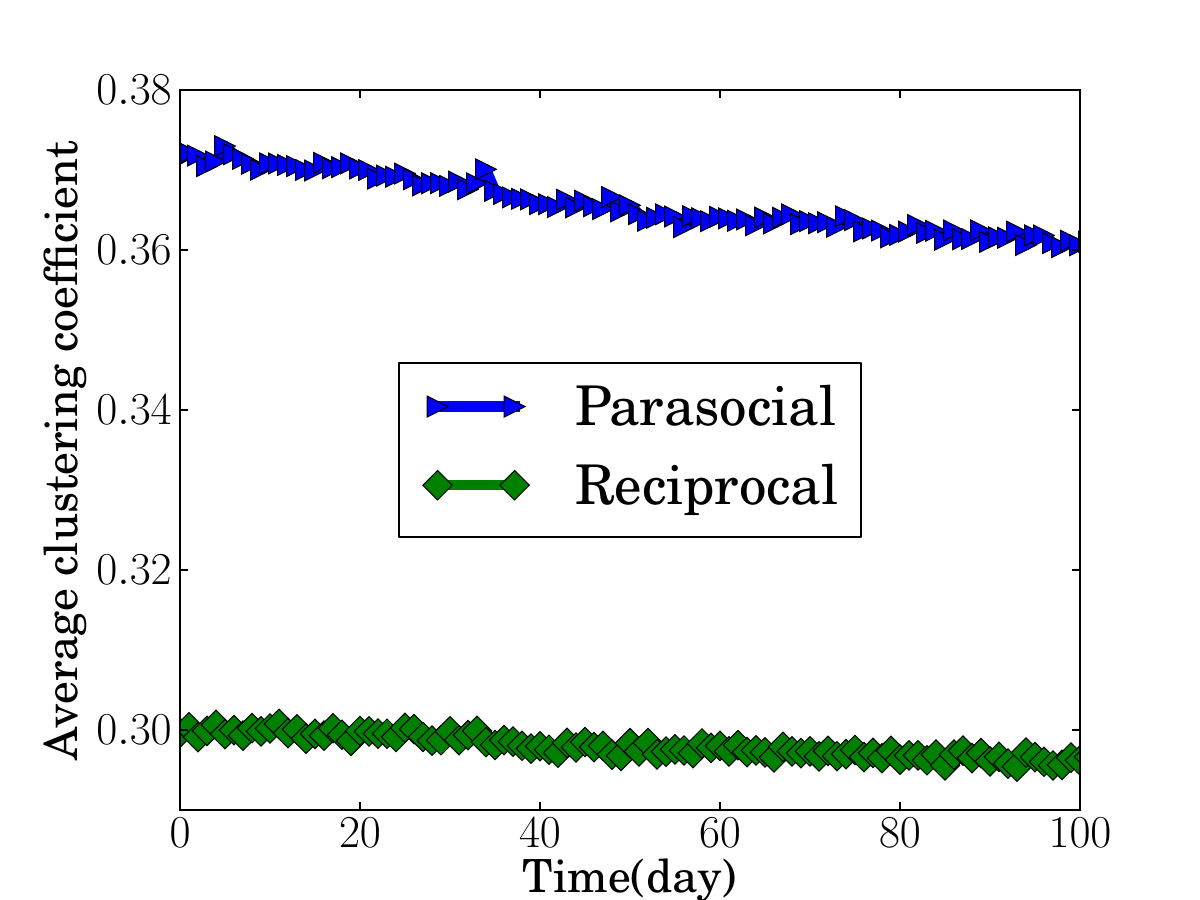} \label{flickr-clustering-evo}}
\caption{Evolution of average clustering coefficients in (a) Google+ and (b) Flickr.}
\label{clustering-evo}
\end{figure*}

\subsection{Global structure and evolution}
\label{sec:assort}
To this end, we look into degree homophily of the parasocial and reciprocal versions of the directed social networks. {Degree homophily characterizes if two linked nodes tend to have similar degrees. Furthermore, assortativity coefficient~\cite{Newman-Assortativity} ranging from -1 to 1 is used to quantify degree homophily.  Intuitively, we can roughly classify users in online social networks into two groups, i.e., \emph{ordinary} users (e.g.,  low-degree users) and \emph{popular} users (e.g., celebrities). Positive assortativity indicates that edges tend to link users within the same group; negative assortativity represents that edges prefer to connect users across the two groups; neutral assortativity means edges link two users without considering which groups they belong to.}

In directed social networks, each node has 4 types of degrees, i.e., outdegree, indegree, parasocial degree, and reciprocal degree. So there are 16 types of degree homophily (e.g., outdegree-indegree homophily, indegree-indegree homophily).  Previous work~\cite{Mislove07,Gong12-imc} explored the outdegree-indegree homophily and found that Flickr has positive assortativity and Google+ has neutral assortativity. However, the outdegree-indegree homophily does not inform us the structure of reciprocal nor parasocial edges. Differently, we study the parasocial-parasocial and reciprocal-reciprocal degree homophily. 

Table~\ref{table:net_stat} shows the assortativity coefficients of parasocial and reciprocal versions of the two online social networks. We observe that they both have \emph{qualitatively} different parasocial and reciprocal assortativities, i.e., parasocial assortativity is neutral while reciprocal assortativity is high positive. This phenomena implies that reciprocal edges are more likely to be intra-group ones (i.e., linking users within the same group) while parasocial edges are more likely to be inter-group ones (i.e., linking users across the two groups). The dominating intra-group reciprocal edges make the reciprocal assortativity high positive. Recall that the parasocial version consists of both the undirected parasocial and reciprocal edges, thus the dominating inter-group parasocial edges neutralize the dominating intra-group reciprocal edges, resulting in a neutral parasocial assortativity. 


Figure~\ref{assort} further illustrates the evolution of the
assortativity coefficients in Google+\footnote{The x-axis of the Google+ evolution figure spans over around 100 days although this Google+ dataset only has 79 daily snapshots because we use the actual crawling date of each snapshot.} and Flickr.  Again, we observe that parasocial and reciprocal assortativity coefficients evolve \emph{qualitatively} differently. Note that reciprocally linking back to a parasocial edge only influences the reciprocal assortativity while parasocial edges only influence the parasocial assortativity. 

Recall that the evolution of Google+ is divided into three phases which is described in Section~\ref{sec:dataset}. Parasocial assortativity keeps stable in Phase I and Phase II and slightly decreases in Phase III. However, reciprocal assortativity increases dramatically in Phase I and decreases in Phase II and Phase III. This implies that new inter-group and intra-group parasocial edges neutralize each other in Phase I and Phase II; after opening to the public (i.e., in Phase III), new inter-group parasocial edges slightly dominate the intra-group ones. Furthermore, new intra-group reciprocal edges significantly dominate inter-group ones in Phase I; new inter-group reciprocal edges slightly dominate the intra-group ones in Phase II and Phase III.  

In Flickr, parasocial assortativity slightly increases while reciprocal assortativity slightly decreases. This implies that new intra-group parasocial edges dominate new inter-group ones in Flickr. Similar to Google+, the decreasing reciprocal assortativity could imply that new inter-group reciprocal edges dominate intra-group ones. 

Both online social networks imply that the impacts of inter-group reciprocal edges on the network structures increase slowly as time evolves. 

%

\subsection{Local structure and evolution}
We study the local structure and evolution of reciprocal and parasocial edges via looking into the clustering coefficients, which characterize how the neighbors of a node are connected.  Given an undirected social network $G=(V,E)$ and a node $u$, $u$'s clustering coefficient is defined as $$c(u) =
\frac{2L(u)}{|\Gamma(u)| (|\Gamma(u)| - 1)}$$, where $L(u)$
is the number of edges among $u$'s neighbors $\Gamma(u)$.
The average \emph{clustering coefficient} is defined as $C=\frac{1}{|V|}\sum_{u\in
V}c(u)$~\cite{Watts98}. Intuitively, this is the average probability that a random pair of neighbors of a random node is connected. 

Previous work~\cite{Newman03,Mislove07,Gong12-imc} found directed social networks have high clustering coefficients. However, these studies cannot demonstrate the local structure of reciprocal and parasocial edges. To this end, we study clustering coefficients of the parasocial and reciprocal versions of social networks.

To determine the clustering coefficient of $u$, we need its degree and the number of edges among its neighbors. On one hand,  parasocial edges increase $u$'s parasocial degree. On the other hand,  parasocial edges also increase the number of edges among $u$'s parasocial neighbors. So one natural question is which one plays a more important role in determining the clustering coefficient. 

Table~\ref{table:net_stat} shows the average clustering coefficients of the parasocial and reciprocal versions of the two social networks. We find that parasocial clustering coefficient is larger than the reciprocal one in both networks. Our observation indicates that parasocial edges, although making nodes' parasocial neighbors more than their reciprocal neighbors, connect the parasocial neighbors more tightly. Note that Cheng et al.~\cite{Cheng11} observed that the reciprocal clustering coefficient is much higher than the parasocial one in a Twitter interaction network, where nodes are Twitter users and a directed edge $(u,v)$ means $u$ has sent some @-messages to $v$. This implies that friendship networks which are our cases are different from the interaction network in terms of the local structure of reciprocal and parasocial edges. 

Figure~\ref{clustering-evo} illustrates evolutions of the clustering coefficients in Google+ and Flickr. We observe that parasocial and reciprocal clustering coefficients evolve in similar patterns. In Google+, both of them decrease dramatically in Phase I and Phase III and increase stably in Phase II. This implies that both of users' parasocial and reciprocal neighbors become more and more loosely connected in Phase I and Phase III while turning to be increasingly tightly connected in Phase II.   In Flickr, both parasocial and reciprocal clustering coefficients decrease over time, which indicates that users in Flickr have increasing number of neighbors and these neighbors are more and more loosely connected. 


\subsection{Summary and implications}

In summary, we find that reciprocal edges are more likely to connect users with similar degrees while parasocial edges are more likely to link ordinary users and popular ones. However, the impacts of reciprocal edges linking ordinary and popular users on the network structures increase slowly  as the social networks evolve. Moreover, parasocial edges, although making nodes' parasocial neighbors more than their reciprocal neighbors, connect the parasocial neighbors more tightly.

Our findings have significant implications for directed social network modeling. Existing directed network models such as Preferential Attachment models~\cite{Barabasi99,Zlati09}, Copying model~\cite{Kleinberg99}, Forest Fire model~\cite{Leskovec05}, Kronecker Graph models~\cite{Leskovec10-JMLR}, and Social-Attribute network model~\cite{Gong12-imc} cannot capture the various degree (i.e., in, out, parasocial, and reciprocal degrees) distributions at the same time. Very recently, Durak et al.~\cite{Durak13} designed a scalable model that can match all specified degree distributions.  Although their work is an important step towards realistic directed social network modeling, it only focuses on degree distributions. 

We observe that the parasocial versions and reciprocal versions of directed social networks behave (qualitatively) differently with respect to various network metrics (e.g., assortativity coefficients and clustering coeffients).  These observations are new dimensions that constrain and guide the designs of more realistic directed social network models.  For instance, one natural question is how to generate a directed social network whose parasocial  version has neutral assortativity but reciprocal version has positive assortativity. Our observations of the global and local structures of the reciprocal and parasocial edges could give us insights on the designs of better directed social network models.


\section{Formation of Reciprocal Edges}
\label{sec:fea}
In this section, we provide a more fine-grained study about the evolution of parasocial and reciprocal edges. Specifically, we find that user behaviors, node attributes, and edge attributes all have impact on the formation of reciprocal edges from parasocial edges. These studies give us insights on how to extract features when predicting reciprocal edges. 

\begin{figure*}[!t]
\centering
\subfloat[Acceptance local reciprocity]{\includegraphics[width=0.33 \textwidth, height=1.8in]{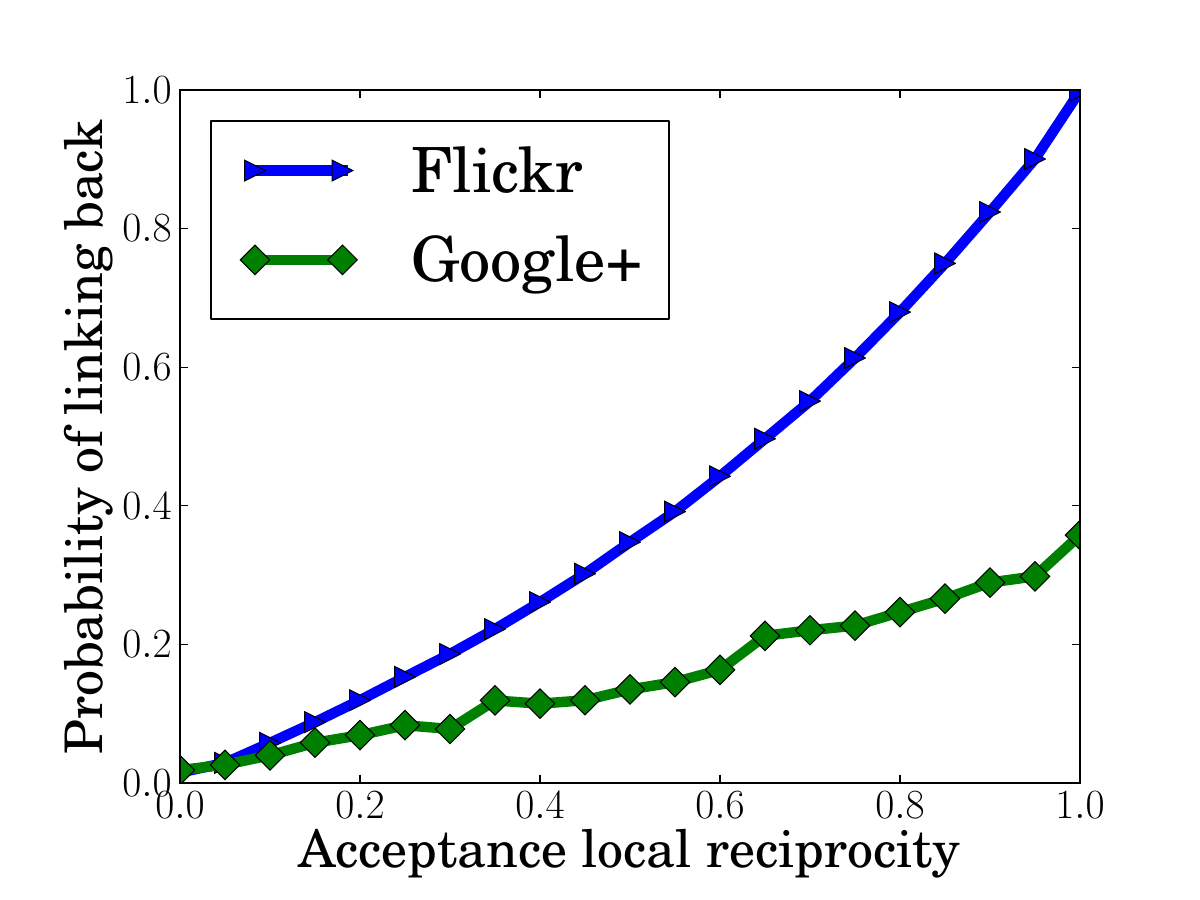}\label{local-reciprocity}}
\subfloat[Node attributes]{\includegraphics[width=0.33 \textwidth, height=1.8in]{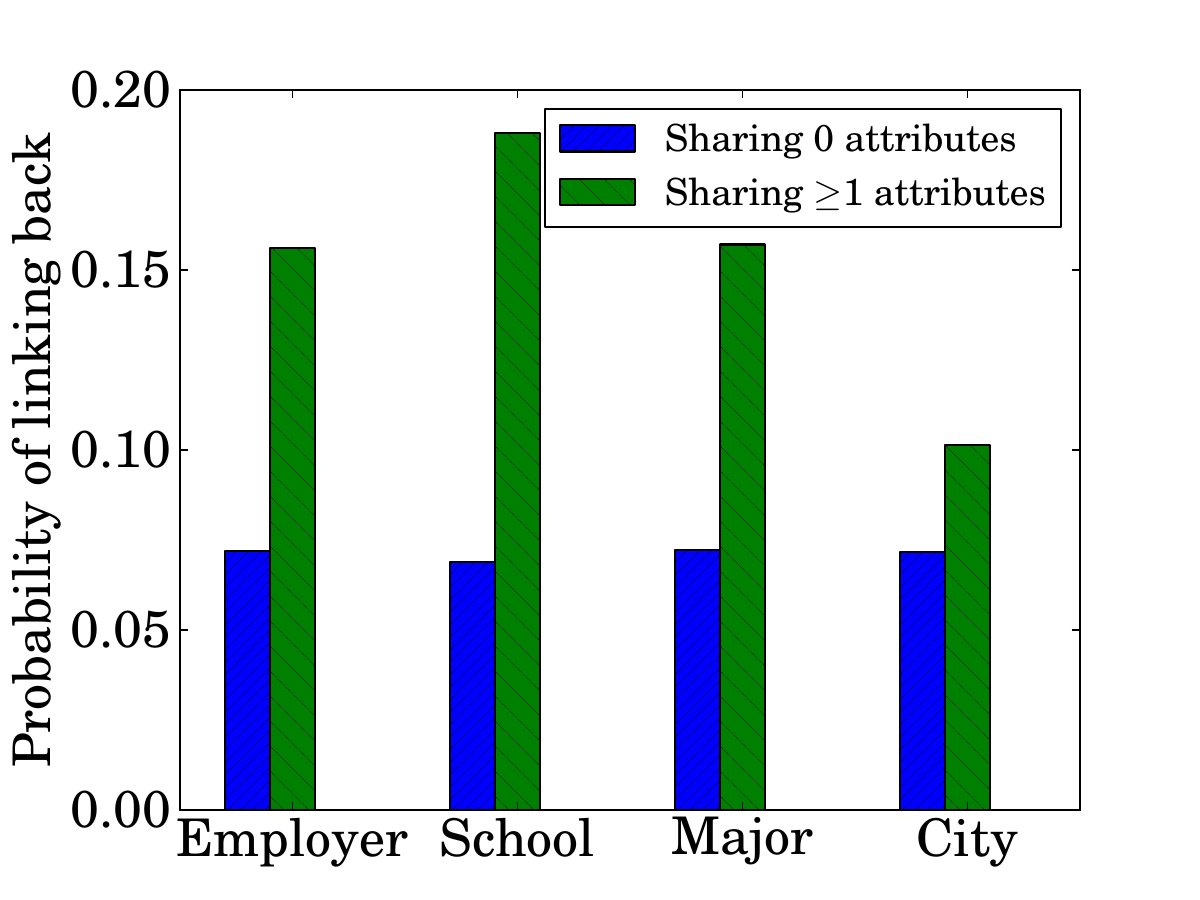}\label{node-attribute}}
\subfloat[Edge age]{\includegraphics[width=0.33 \textwidth, height=1.8in]{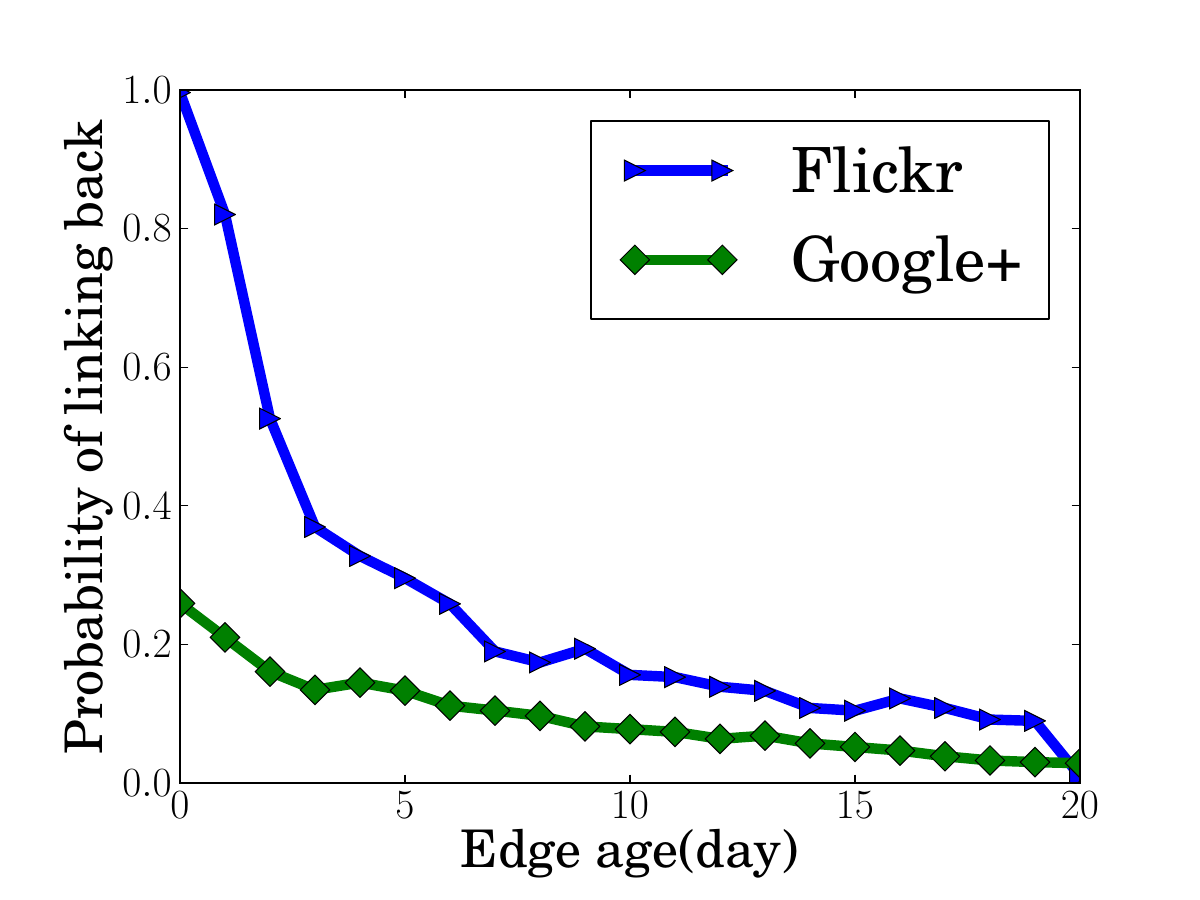}\label{edge-age}}
\caption{The impacts of (a) acceptance local reciprocity, (b) node attributes and (c) edge age on the formation of reciprocal edges. (b) was obtained with only the Google+ dataset since the Flickr dataset doesn't have node attributes.}
\end{figure*}

\subsection{User behavior}
\label{sec:local-reci}
Intuitively, users in online social networks could behave very differently in terms of issuing and accepting friend requests. For instance, one speculation is that users having higher tendencies to accept friend requests in the past are also more likely to accept new friend requests in the future; users whose friend requests were more likely to be accepted in the past will have higher probabilities of their friend requests being accepted in the future. To quantify these intuitions, we define respectively request and acceptance local reciprocity as $R_r(u) = d_r(u)/d_o(u)$ and $R_a(u) = d_r(u)/d_i(u)$. Please refer to Section~\ref{sec:data} for the definitions of $d_o, d_i$ and $d_r$. $R_r(u)$ characterizes the likelihood that $u$'s friend requests are accepted; $R_a(u)$  characterizes the probability with which $u$ accepts a friend request.

To demonstrate the impact of local reciprocities on the formation of reciprocal edges, we calculate the probability of linking back to a parasocial edge as a function of local reciprocities. Specifically, we first discretize the local reciprocity ranging from 0 to 1 to 20 bins. For each bin $b$, we collect all parasocial edges in $\mathrm{G}^{(20)}$ whose tail points have $R_r(u)$ in $b$ or head points have $R_a(u)$ in $b$, and compute the corresponding linking-back probability as the fraction of them that turn to be reciprocal in the last snapshot of Google+ or Flickr.  Figure~\ref{local-reciprocity} shows the linking-back probability as a function of acceptance local reciprocity. We conclude that the linking-back probability increases as acceptance local reciprocity increases. For instance, in Flickr, users that always reciprocally linked back to friend requests (i.e., users with acceptance local reciprocity 1) in the past also always do so in the future. The results of request local reciprocity are similar and thus are not shown here for simplicity. These findings support our speculations and the predictiveness of the local reciprocity for reciprocity prediction.

%

\subsection{Node attributes}
We have node attributes in only Google+ dataset, on which we will focus in this section. Note that 78\% of the Google+ nodes have no available attributes. To avoid the influences of these missing attributes, we use nodes with at least one attribute and edges between them. 

Recall that each node in the Google+ dataset could have four attribute types (i.e., School, Major, Employer and City). Figure~\ref{node-attribute} shows their impacts on the formation of reciprocal edges. The figure was computed as follows: we find all parasocial edges whose end points have at least 1 attribute and share 0 or at least 1 attribute of some attribute type in $G^{(20)}$, then the corresponding linking-back probability is the fraction of such edges that are reciprocal in the last snapshot. 

We observe that the four attribute types have different impacts on the formation of reciprocal edges. For instance, sharing the attribute type \emph{School} triples the linking-back probability while sharing \emph{City} just increases the probability by one third. On one hand, these findings indicate the node attributes are useful for predicting reciprocity. On the other hand, they inform us to consider the four attribute types seperately. Note that Hopcroft et al.~\cite{Hopcroft11} found that sharing the same time zone doesn't increase the linking-back probability in Twitter. However, we observe that sharing cities which are more fine-grained location information does increase the linking-back probability. Gong et al.~\cite{Gong12-imc} showed that sharing common node attributes increases the linking-back probability. However, they didn't distinguish between different attribute types.  Here, we provide a fine-grained study about the impact of different attribute types. 

%


\subsection{Edge attributes}
In general, edge attributes could be messages sent from $u$ to $v$, comments/likes $u$ writes to $v$'s posts/pictures and replying tweets, etc. For instance, Hopcroft et al.~\cite{Hopcroft11} found that retweeting or replying users' tweets increases the probability of linking back to a parasocial edge in Twitter. 

Here, in the Google+ and Flickr datasets, we treat the age of an edge $(u,v)$ as its edge attribute. Similar to the measurement studies on local reciprocity and common attributes, we ensemble all parasocial edges in $\mathrm{G}^{(20)}$ whose ages are $a$, then we compute the linking-back probability with respect to $a$  as the fraction of these edges that become reciprocal in the last Google+ or Flickr snapshot. Since the time resolution of both datasets is a day, we compute the edge ages with respect to days. For instance, edges with age 0 are the new edges in $G^{(20)}$. Figure~\ref{edge-age} shows the linking-back probability decreases dramatically as the edge age increases from 0 to 3 and decreases relatively slowly when the edge age ranges from 4 to 20.  For instance, linking-back probability of parasocial edges with age 0 is around 100 times higher than that of the parasocial edges with age 20 in Flickr.  These results imply that edge age is useful for predicting reciprocity.

\begin{table*}[t]\renewcommand{\arraystretch}{1}
\centering
\caption{Summary of features. Our new features are indicated by the marker *.}
\begin{tabular}{|c|c|c|c|}\hline
\multicolumn{1}{|c|}{Categories} & \multicolumn{1}{|c|}{Names} & \multicolumn{1}{|c|}{Notations} & \multicolumn{1}{|c|}{Definitions}\\ \hline 
\multirow{7}{*}{\small Single-node} & \multicolumn{3}{|c|}{Degrees}\\ \cline{2-4} 
& {\small Indegree} &{\small $d_i(u)$} &{\small $|\Gamma_i(u)|$} \\ \cline{2-4} 
 & {\small Outdegree} &{\small $d_o(u)$} & {\small $|\Gamma_o(u)|$} \\ \cline{2-4} 
 & {\small Outdegree/Indegree ratio} &{\small $d_o(u)/d_i(u)$ } &{\small $|\Gamma_o(u)|/|\Gamma_i(u)|$} \\ \cline{2-4} 

 &\multicolumn{3}{|c|}{\bf Local reciprocities*} \\ \cline{2-4}
&{\small Acceptance reciprocity} &{\small $R_a(u)$ } &{\small $d_r(u)/d_i(u)$}  \\ \cline{2-4} 
&{\small Request reciprocity} &{\small  $R_r(u)$ }&{\small $d_r(u)/d_o(u)$} \\ \hline \hline

\multirow{12}{*}{\small Node-pair} & \multicolumn{3}{|c|}{Structures} \\ \cline{2-4}

& {\small Common neighbors} &{\small $CN_{xy}(u,v)$} & {\small $|\Gamma_x(u)\cap \Gamma_y(v)|$, $x,y \in \{i, o, p, r\}$.} \\ \cline{2-4}
& {\small Jaccards coefficients} &{\small $JC_{xy}(u,v)$} &{\small ${|\Gamma_x(u) \cap \Gamma_y(v) |}/{|\Gamma_x(u) \cup \Gamma_y(v)|}$, $x,y \in \{i, o, p, r\}$} \\ \cline{2-4} 
& {\small Adamic-Adar} &{\small $AA_{xyz}(u,v)$ } &{\small $\sum_{w\in \Gamma_x(u) \cap \Gamma_y(v)} \frac{1}{log(|\Gamma_z(w)|)}$, $x,y,z \in \{i, o, p, r\}$} \\ \cline{2-4} 
& \multirow{2}{*}{\small Preferential attachment} &{\small $d_o(u)\cdot d_i(v)$} & {\small} \\
& &{\small $d_i(u)\cdot d_o(v)$} &  \\ \cline{2-4} 
&\multicolumn{3}{|c|}{\bf Node attributes*} \\ \cline{2-4}

& {\small Common attri. neighbors} &{\small  $\mathrm{CN-A}_a(u, v)$ }&{\small $|\Gamma_a(u) \cap \Gamma_a(v)|$ for four attri. types} \\ \cline{2-4} 
& {\small Attri. Jaccards coefficients} &{\small $\mathrm{JC-A}_a(u,v)$} &{\small $|\Gamma_a(u) \cap \Gamma_a(v)|$/$|\Gamma_a(u) \cup \Gamma_a(v)|$ for four attri. types} \\ \cline{2-4} 
& {\small Attri. Adamic-Adar}&{\small $\mathrm{AA-A}_a(u,v)$}&{\small $\sum_{b\in \Gamma_a(u) \cap \Gamma_a(v)} \frac{1}{log(|\Gamma_s(b)|)}$ for four attri. types} \\ \cline{2-4} 
&\multicolumn{3}{|c|}{\bf Edge attributes*} \\ \cline{2-4}
&{\small Edge age} & {\small $\mathrm{Age}(u, v)$} &{\small Time elapsed since $(u,v)$ exists.}  \\ \hline
 
\end{tabular}
\label{table:features}
\end{table*}

\section{Predicting Reciprocal Edges}
In this section, we study the prediction of reciprocal edges. Given a few social network snapshots, reciprocal edge prediction is to infer which parasocial edges will turn to be reciprocal in the future. 
First, we discuss the extraction of features. Each of the features is supported by either previous work or our measurement studies in Section~\ref{sec:fea}. Then, we map the reciprocal edge prediction to an outlier detection problem.


\subsection{Features}
\label{sec:feature}
We extract two categories of features for each directed edge. The two categories are \emph{single-node features} and \emph{node-pair features}. Table~\ref{table:features} summarizes these features. Our new features are indicated by the star marker * in the table. In the following, we will elaborate them one by one.

\subsubsection{Single-node features} 
These features are extracted for each node individually. For an edge $(u,v)$, the single-node features of both $u$ and $v$ are concatenated.

\myparatight{Degrees~\cite{Cheng11}} Cheng et al. showed that indegree and outdegree and their ratio are useful features for reciprocal edge prediction. Moreover, Hopcroft et al.~\cite{Hopcroft11} also found that high-degree users link back to high-degree users with a higher probability.  
Following Cheng et al., we extract $d_i(u)$, $d_o(u)$, $d_o(u)/d_i(u)$ for  node $u$ as features.

\myparatight{Local reciprocity} We have shown in Section~\ref{sec:local-reci} that local reciprocities impact the formation of reciprocal edges significantly. So we extract both acceptance and request local reciprocities as features.

\subsubsection{Node-pair features}

We extract three categories of node-pair features for each edge $(u,v)$. They are structural features, node-attribute  and edge-attribute features. 

\myparatight{Structural features} We extract a few classical link prediction features~\cite{Liben-Nowell03,Gong11} such as common neighbors, jaccard coefficients, Adamic-Adar scores and preferential attachment as structural features. 


\begin{itemize}
\item {\bf Common neighbors (CN)} In directed social networks, there are four types of neighbors for a node, i.e., incoming, outgoing, parasocial and reciprocal neighbors. Thus, two nodes $u$ and $v$ could have $4\times 4=16$  types of common neighbors. We denote by $CN_{xy}(u, v)$  these 16 types of common neighbors, where $x,y\in \{i, o, p, r\}$. Note that $CN_{oi}(u, v)$ is equivalent to the number of two-step directed paths from $u$ to $v$, which was shown to be useful for reciprocal edge prediction~\cite{Cheng11}. 


\item {\bf Jaccard's coefficients (JC)}~\cite{Salton83} Jaccard's coefficient is a commonly used similarity
metric in information retrieval. The intuition behind the Jaccard's coefficient is to penalize the common social neighbors by the total number of social neighbors the two users have. Formally, $JC_{xy}=\frac{|\Gamma_x(u) \cap \Gamma_y(v) |}{|\Gamma_x(u) \cup \Gamma_y(v)|}$, where $x,y\in \{i, o, p, r\}$. Since each node has 4 kinds of social neighbors, we  have 16 types of Jaccard's coefficients.  

\item {\bf Adamic-Adar (AA)~\cite{aa03}} Intuitively, we want to downweight the importance of neighbors that are social hubs. AA score quantifies this intuition as $$AA_{xyz}(u,v)=\sum_{w\in \Gamma_x(u) \cap \Gamma_y(v)} \frac{1}{log(|\Gamma_z(w)|)}$$, where $x,y,z\in \{i, o, p, r\}$. Totally, we have 64 AA features. 

\item {\bf Preferential attachment (PA)} PA is empirically observed to be a basic mechanism that edge formation follows in various networks~\cite{Gong12-imc,Newman01-pa,Barabasi99,Leskovec08}. As was proposed by Cheng et al.~\cite{Cheng11}, we calculate $PA(u,v)=d_o(u)\cdot d_i(v)$ and $PA(v,u)=d_i(u)\cdot d_o(v)$ as features. 

\end{itemize}

Note that Cheng et al.~\cite{Cheng11} only extracted $CN_{ii}$, $CN_{oo}$, $JC_{ii}$, $JC_{oo}$, $AA_{iii}$ and PA as structural features.  Hopcroft et al.~\cite{Hopcroft11} used $CN_{rr}$ as features. 

\myparatight{Node-attribute features} We have node attributes in only Google+ dataset, on which we will focus in this section. However, we want to stress that our ways of extracting node attributes features can be naturally generalized to other social networks. 
\begin{itemize}
\item{\bf Common attribute neighbors (CN-A)} Recall that we have shown in Figure~\ref{node-attribute} that the four attribute types (i.e., School, Major, Employer and City) have different impacts on the formation of reciprocal edges. So we consider them seperately. Specifically, we extract the number of common attribute neighbors for each attribute type as features.  

\item {\bf Attribute Jaccard's coefficients (JC-A)/Adamic-Adar (AA-A)} Similar to structural JC and AA features, we downweight the importance of a common attribute with either the total number of  attributes the two users have or the number of social neighbors of the attribute. Formally, attribute JC is defined as $\mathrm{JC-A}_a(u,v)$ = $\frac{|\Gamma_a(u) \cap \Gamma_a(v)|}{|\Gamma_a(u) \cup \Gamma_a(v)|}$; attribute AA is defined as $$\mathrm{AA-A}_a(u,v)=\sum_{b\in \Gamma_a(u) \cap \Gamma_a(v)} \frac{1}{log(|\Gamma_s(b)|)}$$. Again, we extract these features for the four attribute types seperately.
\end{itemize}

\myparatight{Edge-attribute features} We have shown in Figure~\ref{edge-age} that edge age impacts the formation of reciprocal edges significantly. So we extract a feature from the edge age. For a parasocial edge, this feature is simply its age. However, it's trickier to extract this feature for a reciprocal edge.  Note that most of the reciprocal edges could have large ages since they might appear a long time ago, making them indistinguishable with large-age parasocial edges. However, a reciprocal edge might have a small age at the time when it became reciprocal. So we extract this age as the feature.

\subsection{Modeling reciprocal edge prediction as outlier detection}
Previous studies treat reciprocal edge prediction as either a supervised~\cite{Cheng11} or a semi-supervised learning problem~\cite{Hopcroft11}. With snapshot $G^{(t)}$, their approaches treat reciprocal edges as positive examples and sample some parasocial edges as negative examples when training the models. However,  the parasocial edges in $G^{(t)}$ are also test examples in $G^{(t+1)}$ and some of the sampled ones might have become positive in $G^{(t+1)}$. Figure~\ref{outlier}  demonstrates such an issue. In $G^{(t)}$, edge $(w,u)$ is a positive example, and edge $(v,u)$ is sampled as a negative example. However, $(v,u)$ turns to be positive in the test snapshot $G^{(t+1)}$. As a result, the better their trained models are, the worse generalization performances they possibly achieve. 

Actually, in the setting of reciprocal edge prediction, we can only observe positive examples (i.e., reciprocal edges). So we propose to model reciprocal edge prediction as an outlier detection problem with known positive examples.  We'll use one-class Support Vector Machine~\cite{Manevitz01} as an outlier detector.

\section{Evaluations}
\label{sec:eva}
We show our experimental results of reciprocal edge prediction. First, we introduce our experimental setup. Then we compare our proposal with previous approaches. 

\subsection{Experimental setup} In the following, we will cover the construction of the training and test datasets, approaches we compare our proposal with, data normalization techniques we apply to the feature matrix, and metrics adopted.

\myparatight{Training and test} 
According to the definition, reciprocal edges exist as pairs of edges. For instance, both $(w,u)$ and $(u,w)$ in Figure~\ref{outlier} are reciprocal edges. However, to make our description more clear, for a pair of reciprocal edges, we only call the one that appeared earlier as a reciprocal edge in the entire Section~\ref{sec:eva}. To continue the above example, only $(w,u)$ is called reciprocal since it appeared earlier than $(u,w)$. Intuitively, these newly defined reciprocal edges are friend requests that are already accepted.  However, we keep the original definition of parasocial edges.

Around 78\% of users have no available node attributes in the Google+ dataset. Thus, in order to avoid the influences of the missing attributes, we further preprocess Google+ via only keeping nodes with at least one attribute and edges between them. However, we should point out that the features of an edge are still extracted from the original Google+. 

We use $\mathrm{G}^{(20)}$ of Flickr and the preprocessed Google+ as the training dataset. Note that the parasocial edges in $\mathrm{G}^{(20)}$ are also the test dataset. We choose $G^{(20)}$ to construct the training and test datasets because we can extract edge age features for the edges in them and the parasocial edges  have enough time to become reciprocal in the last snapshots if they would be.

\begin{table}[!t] \renewcommand{\arraystretch}{1}
\centering
\caption{Statistics of the training and test datasets.}
\addtolength{\tabcolsep}{-2pt}
\begin{tabular}{|c|c|c|c|c|}\hline
&\multicolumn{2}{|c|}{\small Training} & \multicolumn{2}{|c|}{\small Test} \\ \cline{2-5}
&{\small Reciprocal} & {\small Parasocial} &{\small Reciprocal} & {\small Parasocial}\\ \hline
{\small Google+} &{\small 11,993,458} & {\small 16,250,289} & {\small 1,186,903} & {\small 15,063,386}\\ \hline 
{\small Flickr} &{\small 8,211,426} &{\small 9,906,869} &{\small 139,762} &{\small 9,767,107} \\ \hline
\end{tabular}
\label{exp-data}
\end{table}

In the training phase, we treat the reciprocal edges in $\mathrm{G}^{(20)}$ as positive training examples. These positive examples are enough to train the outlier detection model. However, supervised and semi-supervised learning models also require negative examples. So, as was done in~\cite{Cheng11,Hopcroft11},  we sample $\alpha P$ parasocial edges in $\mathrm{G}^{(20)}$ and treat them as negative examples, where $P$ is the number of positive training examples. Note that some of the sampled training negative examples are also test positive examples, possibly resulting in bad generalization performances. So we design two sampling strategies, i.e., \emph{random} and \emph{edge-age} sampling.  Random sampling means we sample the $\alpha P$ negative examples uniformly at random. Recall that Figure~\ref{edge-age} shows parasocial edges with larger edge ages are less likely to be reciprocal in the future. So, in order to reduce the number of sampled training negative examples that are also test positive examples, we design edge-age sampling, which samples the $\alpha P$ negative examples with the largest edge ages. We use 2-fold cross validation and grid search to learn the model hyperparameters. Since this procesure is time-consuming, we perform it with 10\% of the training examples sampled uniformly at random. Then we retrain the models on the full training data with the learned hyperparameters. In the test phase, a parasocial edge in $\mathrm{G}^{(20)}$ is a test positive example if it's reciprocal in the last snapshot of Google+ or Flickr, otherwise it's a test negative example. 

Table~\ref{exp-data} shows the statistics of the training and test datasets. 7.3\% and 1.4\% of the training parasocial edges turn to be test reciprocal ones in Google+ and Flickr respectively. These edges make the supervised and semi-supervised learning approaches achieve bad test performances. We note that 9.9\% of the parasocial edges in $G^{(20)}$ turn to be test reciprocal ones in the original Google+ dataset.

\begin{figure*}[!t]
\centering
\subfloat[Precision in Google+]{\includegraphics[width=0.33 \textwidth, height=1.8in]{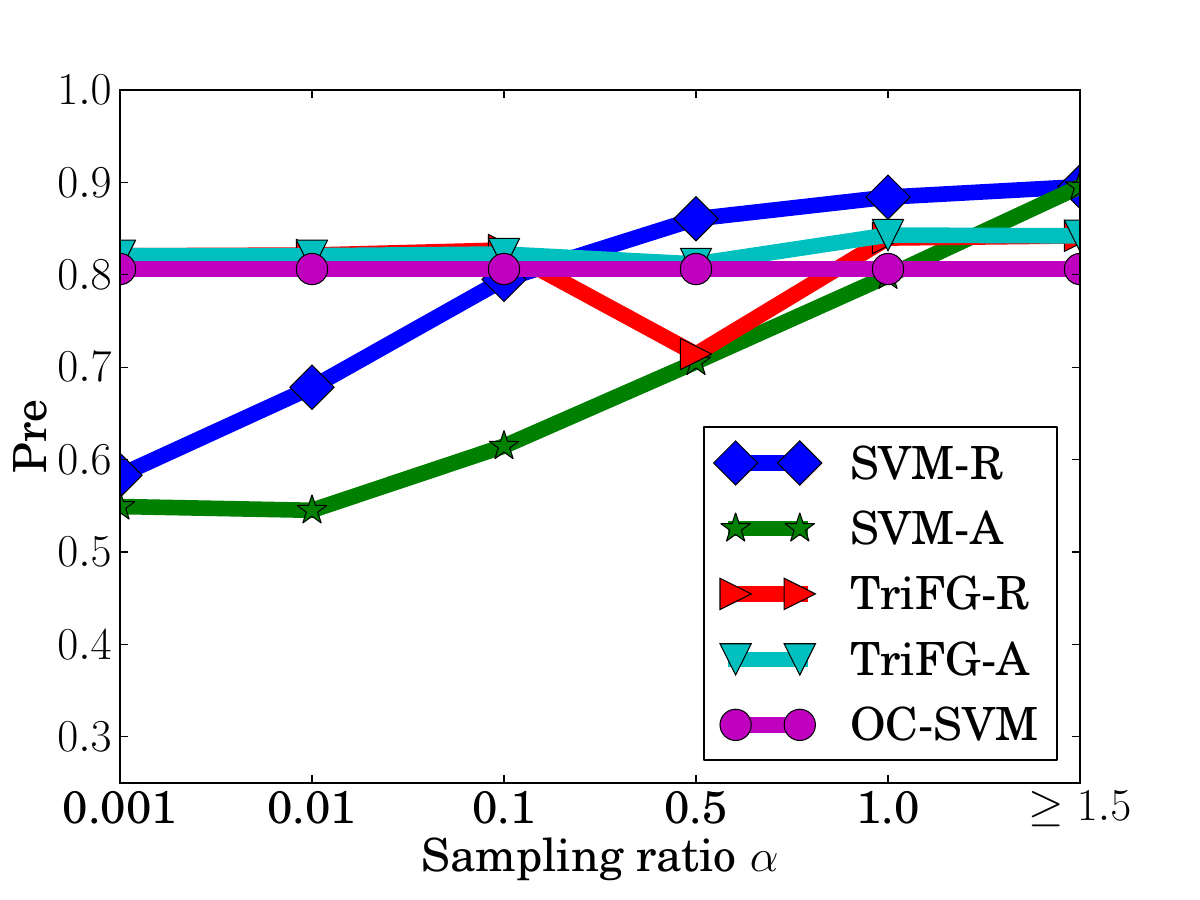} \label{gplus-pre-test}}
\subfloat[Recall in Google+]{\includegraphics[width=0.33 \textwidth, height=1.8in]{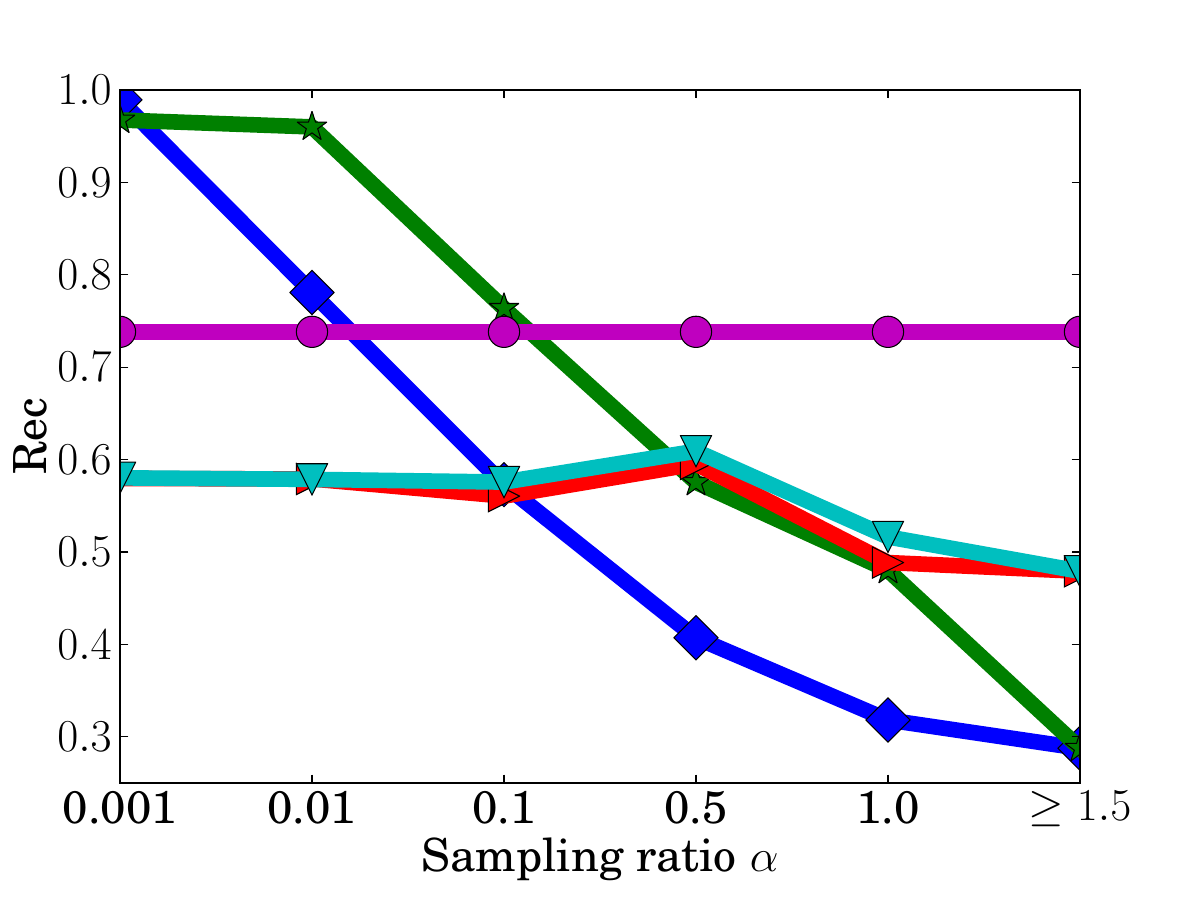} \label{gplus-rec-test}}
\subfloat[F1 in Google+]{\includegraphics[width=0.33\textwidth, height=1.8in]{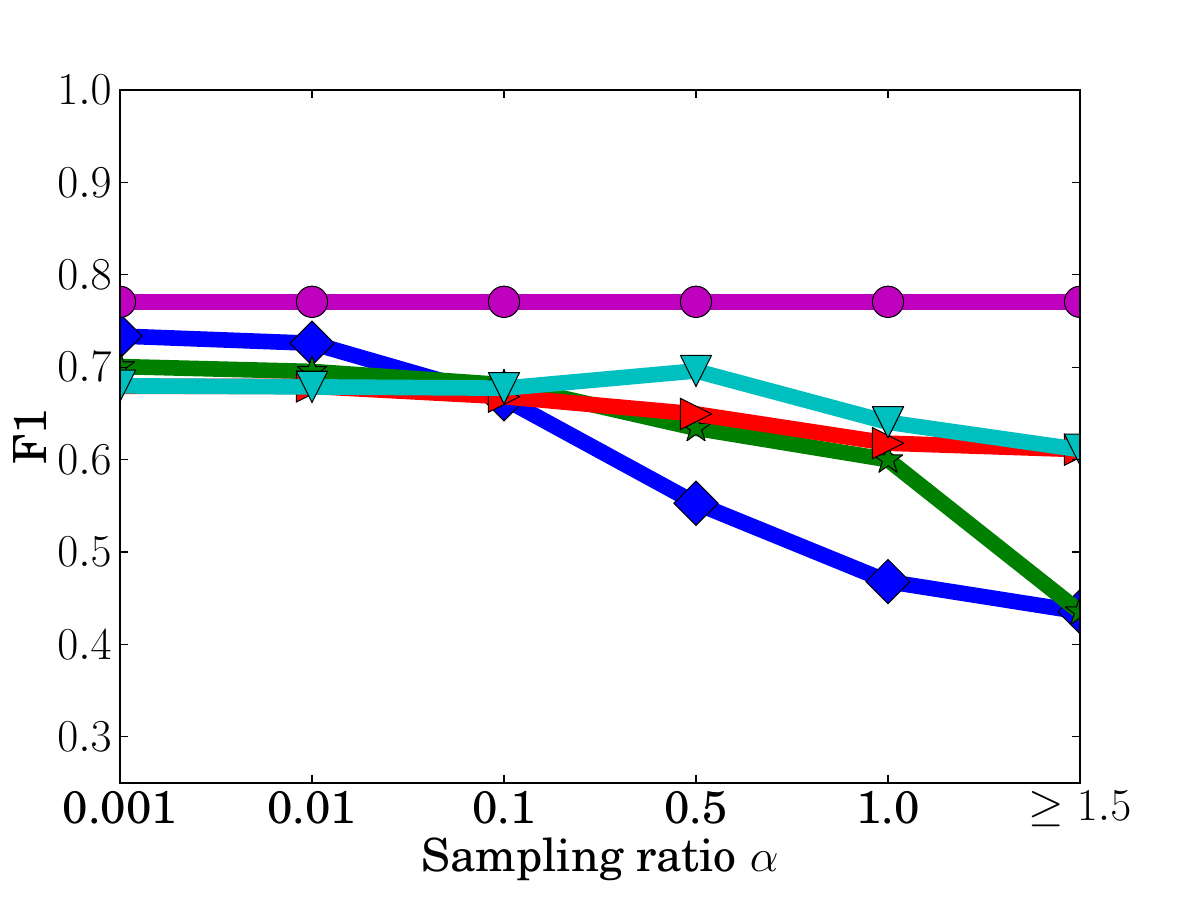} \label{gplus-f1-test}}

\subfloat[Precision in Flickr]{\includegraphics[width=0.33 \textwidth, height=1.8in]{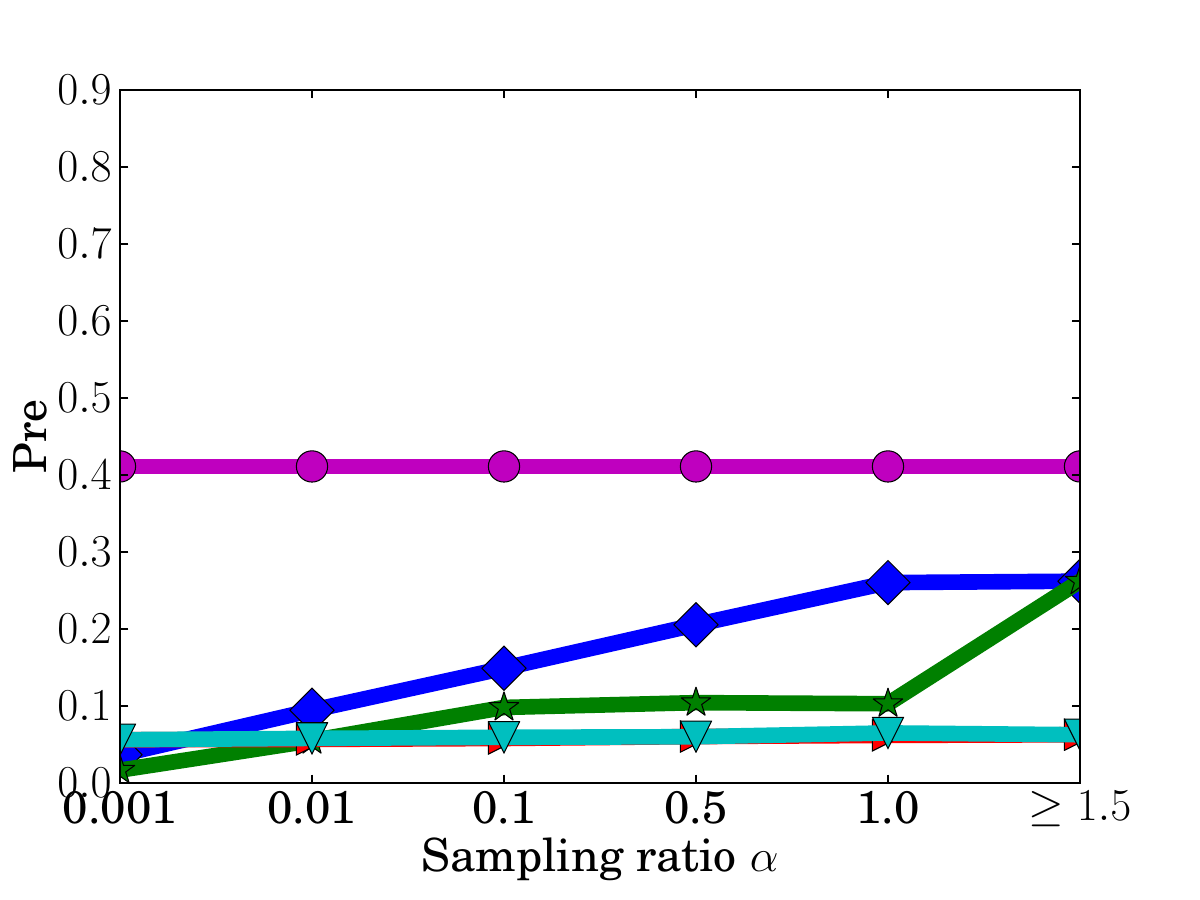} \label{flickr-pre-test}}
\subfloat[Recall in Flickr]{\includegraphics[width=0.33 \textwidth, height=1.8in]{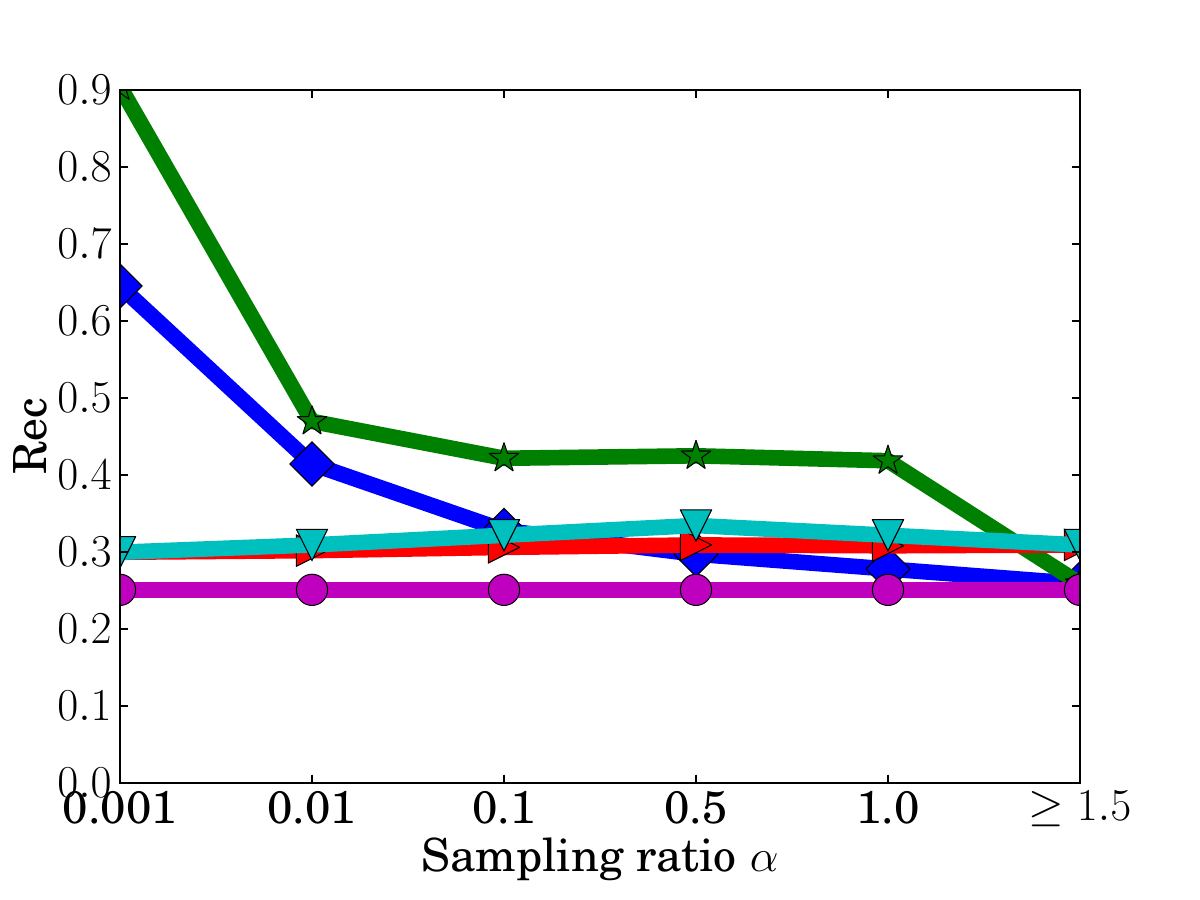} \label{flickr-rec-test}}
\subfloat[F1 in Flickr]{\includegraphics[width=0.33\textwidth, height=1.8in]{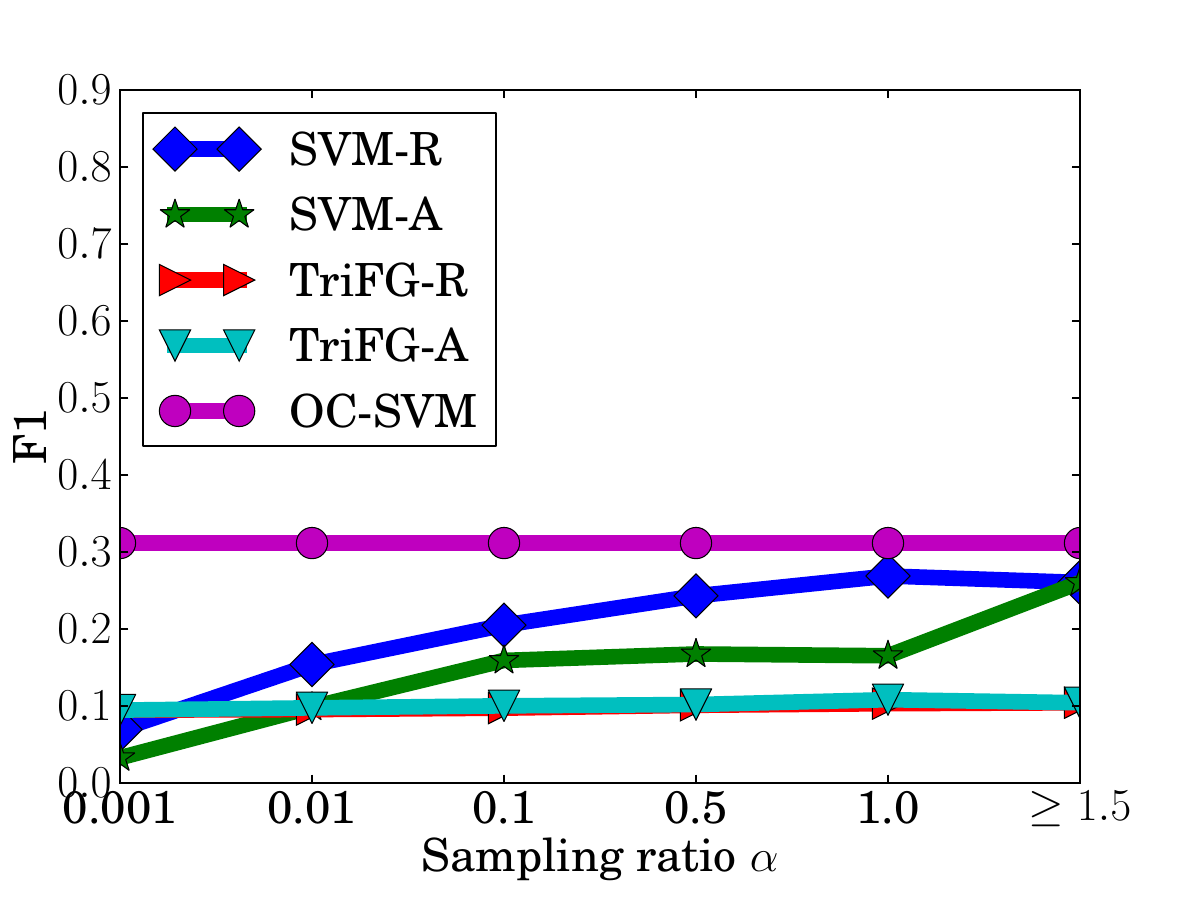} \label{flickr-f1-test}}

\caption{Comparisons between our proposal and previous approaches in the Google+ and Flickr social networks.}
\label{result}
\end{figure*}

\myparatight{Comparisons} We compare the following approaches. 

\begin{itemize}
\item {\bf SVM-R} Binary SVM~\cite{Cortes95} with the random sampling strategy to sample the negative examples. 
\item{\bf SVM-A} Binary SVM with the edge-age sampling strategy. 
\item {\bf TriFG-R} TriFG~\cite{Hopcroft11} is a semi-supervised learning framework based on a factor graphical model. Apart from all the features discussed in Section~\ref{sec:feature}, TriFG also incorporates \emph{structural balance}~\cite{Cartwright56} by modeling it as a factor in the factor graphical model. So we also extract structural balance features as was done in Hopcroft~\cite{Hopcroft11} when testing TriFG. TriFG-R samples the negative examples with  the random sampling strategy.
\item {\bf TriFG-A} TriFG with the edge-age sampling strategy.
\item {\bf OC-SVM} One-class Support Vector Machine~\cite{Manevitz01} is an outlier detector with known positive examples. Note that we don't need to sample negative examples for OC-SVM.
\end{itemize}

We only consider linear kernels for SVM and OC-SVM due to scalability issues. We use  LIBSVM\footnote{{http://www.csie.ntu.edu.tw/\url{~}cjlin/libsvm/}} for OC-SVM implementation and LIBLINEAR\footnote{http://www.csie.ntu.edu.tw/\url{~}cjlin/liblinear/} for SVM implement. LIBLINEAR's linear SVM implementation is much faster than LIBSVM's. However, LIBLINEAR doesn't have OC-SVM implementation for now. TriFG implementation was obtained from the authors~\cite{Hopcroft11}. All these algorithms were run on a machine with 500GB main memory and 32 cores.

\myparatight{Data normalization} We ensemble all the feature vectors into a feature matrix,  rows of which correspond to edges and columns of which correspond to features. It's well known that the performances of many machine learning algorithms are sensitive to data normalizations. So we apply three normalization techniques to this feature matrix. They are i) \emph{column normalization}, which normalizes each column of the feature matrix to have mean 0 and variance 1, ii)  \emph{row normaliztion}, which  normalizes each row to have L2 norm 1, and iii) \emph{column-row normaliztion}, which sequentially applies column and row normalizations to the feature matrix.

We find that algorithms tested in the following perform the best with different normalization techniques. However, we will only show the results with the best normalization for simplicity.

\myparatight{Metric} The number of reciprocal and parasocial edges are highly imbalanced in the test phase. Thus accuracy is not an appropriate metric. For instance, a naive model which always outputs negative can already achieve test accuracy 0.986 in Flickr and 0.927 in Google+. So, as was done in Hopcroft et al.~\cite{Hopcroft11}, we adopt Precision, Recall and F1 as the metrics. Precision is the portion of predicted positive examples that are true reciprocal edges. Recall is the portion of true reciprocal edges that are predicted as positive. F1 score is the harmonic mean of Precision and Recall.

\subsection{Comparison results}
Figure~\ref{result} shows the test performances of the approaches as functions of the sampling ratio $\alpha$. $\alpha \geq 1.5$ corresponds to the scenario in which all parasocial edges are sampled as negative examples. We have the following observations.

\myparatight{Google+ vs. Flickr} Google+ and Flickr achieve quantitatively different performances. Overall, it is easier to predict reciprocal edges in Google+ than in Flickr. For instance, Google+ achieves around 0.8 Precision while Flickr achieves around 0.4 Precision for OC-SVM. Moreover, they behave qualitatively differently with respect to F1 scores. Specifically, F1 score decreases in Google+ while it increases in Flickr as the sampling ratio $\alpha$ increases for all approaches except OC-SVM. We find that one reason is that the Google+ dataset has node attributes while the Flickr dataset does not\footnote{Our experimental results show that performances in Google+ decrease if we do not use features related to node attributes. We do not show the corresponding results for brevity.}. We speculate another reason is that the Google+ dataset represents an OSN's early stage while the Flickr dataset represents an OSN's steady stage.

\myparatight{OC-SVM vs. SVM and TriFG} OC-SVM achieves better F1 scores than SVM and TriFG approaches in both Google+ and Flickr datasets. On one hand, SVM and TriFG approaches sample some parasocial edges as negative examples in the training phase. However, around 7.3\%  and 1.4\% of them are also test positive examples in Google+ and Flickr respectively, which cause bad test performances. On the other hand, OC-SVM doesn't use the parasocial edges in the training phase, thus avoids this issue.   


\myparatight{Random sampling vs. edge-age sampling} Edge-age sampling achieves better Recall than random sampling for both SVM and TriFG approaches. We take SVM with the sampling ratio $\alpha=1.0$ and Google+ as an example to illustrate the reason. Specifically, 875,989 sampled negative examples are actually test positive examples with random sampling and 580,726 of them are classified as negative in the training and test phases. However, the number of such examples decreases to 567,620 with edge-age sampling and 440,346 of them are classified as negative. The behavior of Precision is more complicated. Specifically, edge-age sampling slightly helps TriFG but makes SVM perform worse. The reason might be that the subset of parasocial edges sampled by edge-age sampling are biased to have large edge ages, which decreases SVM-A's Precision.  

\myparatight{Impact of the sampling ratio} For SVM approaches, Precision increases and Recall decreases as the sampling ratio $\alpha$ increases. On one hand, a larger $\alpha$ treats more parasocial edges as training negative examples and thus classifies more test reciprocal edges as negative, which explains the decreasing Recall. On the other hand, a larger $\alpha$ also correctly classifies more test parasocial edges as negative and thus decreases the number of test parasocial edges among the predicted reciprocal edges, which increases the Precision. However, the behavior of the F1 score depends on social networks. Specifically, SVM's performances increase in Google+ but decrease in Flickr when $\alpha$ goes to 0. Interestingly, TriFG approaches are relatively robust to the selection of $\alpha$ with respect to all the three metrics. We speculate the reason is that TriFG incorporates structural balance information.

\section{Related Work}
We first briefly review previous measurement studies involving reciprocity, then we review approaches to predict reciprocal edges. Finally, we discuss the differences between reciprocal edge prediction and other classical link mining tasks.

\myparatight{Measuring reciprocity}  A few studies looked into the reciprocity (i.e., the fraction of links that are symmetric) in a static snapshot of an OSN. For instance, Mislove et al.~\cite{Mislove07} measured the reciprocity to be 0.62 on Flickr and 0.79 on YouTube, and Kwak et al.~\cite{Kwak10} measured the reciprocity to be 0.22 on Twitter. More recently, Gong et al.~\cite{Gong12-imc} explored the evolution of reciprocity and found that Google+'s reciprocity decreases as Google+ evolves.

Zlati et al.~\cite{Zlati09} and Lopez et al.~\cite{L08pez} discussed the influence of reciprocity on degree correlations. Akoglu et al.~\cite{Akoglu12} quantified reciprocity in weighted communication networks (e.g., phone call network). Singhal et al.~\cite{Singhal13} studied reciprocity and its evolution in three related \emph{interaction} networks (i.e.,  chat, trade, and trust networks) which describe behaviors of users in an online game. They also found that features extracted from multiple networks can enhance the prediction accuracy of reciprocity.  Seshadhri et al.~\cite{Seshadhri13} proposed a collection of directed closure metrics for directed networks with reciprocal edges, and they found that reciprocal edges significantly influence the formation of triangles.

Cheng et al.~\cite{Cheng11} found that reciprocal version has a higher average clustering coefficient than parasocial version in their Twitter interaction network, which implies that friendship networks (e.g., Google+ and Flickr) and the interaction network have different local structures of reciprocal and parasocial edges. Hopcroft et al.~\cite{Hopcroft11} found that, in a Twitter follower network, network structure, users' social status and interactions influence the formation of reciprocal edges. 


\myparatight{Predicting reciprocal edges} Cheng et al.~\cite{Cheng11} treated reciprocal edge prediction as a supervised learning problem. Hopcroft et al.~\cite{Hopcroft11} constructed a semi-supervised learning framework for reciprocal edge prediction. These approaches require negative training examples, which are sampled from parasocial edges. Unfortunately, these sampled negative examples are also test examples, and some of them will turn to be positive in the test phase, which possibly decreases the generalization performances. On the contrary, we identify the reciprocal edge prediction is better modeled as an outlier detection problem.    

\myparatight{Differences with other link mining tasks} reciprocal edge prediction is related to a few other link mining tasks such as \emph{link prediction} and \emph{link sign prediction}.

The link prediction problem aims to identify links that are missing in the current  network snapshot but are possible to appear in the near future~\cite{Liben-Nowell03,Gong11,Sarkar12,Sarkar11,Hasan06}. reciprocal edge prediction and link prediction differ in several important aspects. First, previous work~\cite{Cheng11} has shown that features working well for link prediction are not the most effective ones for reciprocal edge prediction. Second, in the setting of reciprocal edge prediction, a parasocial edge already exists between two nodes, from which we can extract features (e.g., edge age).
 Third, they have very different domains. Specifically, the domain of the link prediction problem includes all  non-existing links in the network, which is huge. However, the domain of the reciprocal edge prediction problem consists of the parasocial links, which is much smaller. The domain sizes have significant impact on the efficiency of the corresponding algorithms. 

The link sign prediction~\cite{Leskovec10} classifies social relationships to be either positive (e.g., friendship) or negative (e.g., opposition or antagonism).  This forms interesting contrasts with reciprocal edge prediction in the sense that reciprocal and parasocial links can easily exhibit either type of sign.


\section{Conclusion and Future Work}
In this paper, we first compare the structures and evolutions of reciprocal edges with those of parasocial edges in Google+ and Flickr. We find that reciprocal edges are more likely to connect users with similar degrees while parasocial edges are more likely to link \emph{ordinary} users (e.g., low-degree users) and \emph{popular} users (e.g., celebrities). However, the impacts of reciprocal edges linking ordinary and popular users on the network structures increase slowly as the social networks evolve. Moreover, parasocial edges, although making nodes' parasocial neighbors more than their reciprocal neighbors, connect the parasocial neighbors more tightly. Second, we find that user behaviors, node attributes, and edge attributes all have significant influences on the formation of reciprocal edges. Third, in contrast to previous studies that treat reciprocal edge prediction as either a supervised or a semi-supervised learning problem, we identify that reciprocal edge prediction is better modeled as an outlier detection problem. Finally, we perform extensive evaluations with the Google+ and Flickr datasets, and we demonstrate that our proposal outperforms previous ones.

A few interesting future work includes designing a directed network model that matches real networks with respect to not only directed network metrics but also undirected ones of the corresponding parasocial and reciprocal versions,  designing an outlier detecter incorporating the structural balance features, {and exploring applications of our results in  discovering  users' lifestyles~\cite{Yuan13} and linking users across communities and multiple online social networks~\cite{Liu13}.}

\section{ACKNOWLEDGMENTS}

We thank Dawn Song for her insightful discussions. An initial draft of this paper was available as a technical report~\cite{gongReciprocity}. This work is supported by Intel through the ISTC for Secure Computing, and by a grant from the Amazon Web Services in Education program. Any opinions, findings, and conclusions or recommendations expressed in this material are those of the author(s) and do not necessarily reflect the views of the funding agencies.


%

\end{document}